\documentclass[aps,epsfig,showpacs]{revtex4-1}
\usepackage{amssymb}
\usepackage{graphics}
\usepackage{graphicx}
\usepackage{epstopdf}
\usepackage{float}
\begin{document}
\title{Melting of DNA in confined geometry}
\author{Arghya Maity}
\email{p2014424@pilani.bits-pilani.ac.in}
\author{Navin Singh}
\email{navin@pilani.bits-pilani.ac.in}
\affiliation{Department of Physics, BITS-Pilani, Pilani campus, India-333031}
\begin{abstract}
The stability of DNA molecule during the encapsulation process is a topic of intense research. We study the thermal stability of the double-stranded DNA molecule of different lengths in a confined space. Using a statistical model we evaluate the melting profile of DNA of different length in two geometries: conical and cylindrical. Our results show that not only the confinement but also the geometry of the confined space plays a prominent role in the stability and opening manner of the molecule.

\end{abstract}

\maketitle

\section{Introduction}
\label{intro}
Deoxyribonucleic acid (DNA) is one of the interesting and complex biomolecules.
The most abundant conformation of this molecule is a double-helical structure (also known as the Watson-Crick double helix).
The genetic information of the entire organism is coded in the sequence of four nucleotides {\it Adenine (A)}, {\it Guanine (G)}, {\it Cytosine (C)}, {\it Thymine (T)} \cite{watson}. It has been observed that on increasing the temperature of the solution containing a DNA molecule, the double-strand takes a conformational change into the single-stranded configuration. This is known as {\it DNA melting} or {\it denaturation} \cite{zhang1997,williams, Vologodskii,maxim2014,poland_1966,Fisher1984,Richard2004}. The same can be achieved by changing the pH of the solution as well as by pulling either of the strands by a mechanical device \cite{somen_IJP_2014,hatch_2008,kumar2010}.
The importance of this molecule lies in its wide applications in the molecular motor, DNA computer, origami formation, DNA chip, and biomedicine, etc.  To perform gene therapy \cite{Putam_Nature-2006}, nanorobotics \cite{yurke-nature-2000}, diagnostics \cite{Dharmadi-2004} performance, DNA degradation is a major obstacle for the effective and efficient use of DNA. DNA degradation eventuates in chemical breakdown \cite{Putam_Nature-2006} or sometimes through mechanical forces \cite{murphy-2006}. In gene therapy technique DNA is protected by a physical barrier. Many different techniques have to get better results like complexation with polycations   \cite{Putam_Nature-2006}, charged copolymers of different architecture, cationic lipids or liposomes \cite{maria-2001}. As other options, DNA can be confined within gel   \cite{goh-2004,Anne-2009}, polymeric nanocapsules(micelles)   \cite{Soft_Mat-2012,csaba-2005}, microparticles. There are many elegant and versatile approach for DNA encapsulation \cite{Soft_Mat-20111}.

In DNA encapsulation, the ability to preserve the DNA and efficiently release it, are the processes which are inversely related. The objective of good DNA encapsulation is to find an optimal balance between these two issues. Researchers are trying to balance it in numerous ways; one of them is short DNA encapsulated by a spherical inorganic nanoshell with an overall thickness of $\sim$10 nm \cite{jacs-2010}. Carbon nanotubes also have been proved itself as a potential candidate for DNA encapsulation \cite{Mota_JCP14}. The thermodynamical spontaneity of DNA encapsulation of carbon nanotube under different conditions is still a significant area of research. The threshold diameter of this tube is also a vital issue to investigate since below the threshold encapsulation is inhibited \cite{Mota_BEJ15}. Many sensitive parameters are involved in DNA encapsulation techniques like the medium and topology of the carrier, thermodynamic parameters, etc. 

{\it In vivo}, the DNA molecule is confined in a limited space such as the cell chamber or a channel and is in highly dense solvent conditions \cite{Huaping_jcp14,Sanjay2017,turner,amar_pccp17}. This confinement restricts the conformation and movement of DNA molecules in the cell. The thermodynamic properties of DNA molecules highly depend on the confined space \cite{akabayov}. It is known that conformational properties of biopolymers under confinement are the crucial relevance in living systems like DNA packing in eukaryotic chromosomes, viral capsids, etc. \cite{kumar}. 

In most of the studies, either rectangular or spherical geometry \cite{Huaping_jcp14} is considered as confinement. Despite recent progress, still little is known about the confinement nature of DNA \cite{Phys_Life_Rev_2012_wanunu}. In the present manuscript, motivated by all these different kinds of experiments in different geometries,  we attempt to understand the thermal stability of a DNA molecule that is confined in different geometries through a statistical model. The present manuscript is an following work of our recent published research work \cite{Maity_2019}. It helps to understand confinement effect more immensely. The present work is divided into the following sections: with a brief introduction to the statistical model in Sec. \ref{model}, we discuss the melting of DNA in confined geometry in Sec. \ref{results}. The paper ends with a brief discussion and future scope of the work in Sec. \ref{conc}.

\section{Model and Methods} 
\label{model}

To investigate the effect of confinement on the thermal denaturation of DNA molecule, we adopt the well known Peyrard-Bishop-Dauxois model (PBD). The model is quasi-one-dimensional and describes the motion of the molecules through the stretching of the hydrogen bonds between the bases in a pair \cite{pb,pb1}. The model underestimates the entropy associated with the different conformations of the molecule \cite{frank}. In the linear form of the model, it ignores the effect of helicoidal nature of the molecule as well as the solvent effect of the solution. Despite of these shortcomings, the model still has enough details to describe the denaturation/unzipping process of DNA molecule. The interactions in the DNA, containing $N$ base pairs, are represented as,
\begin{equation}
\label{eqn1}
H = \sum_{i=1}^N\left[\frac{p_i^2}{2m}+ V_M(y_i) \right] + \sum_{i=1}^{N-1}\left[W_S(y_i,y_{i+1})\right],
\end{equation}
here $y_i$ represents the stretching from the equilibrium position of the hydrogen bonds. First term of the model is the momentum term which is $p_i = m\dot{y}_i$. The reduced mass, $m$, is considered the same for both $AT$ and $GC$ base pairs. The interaction between the nearest base pairs along the chain, the stacking interaction, is represented by, 
\begin{equation}
\label{eqn2}
 W_S(y_i,y_{i+1}) = \frac{k}{2}(y_i - y_{i+1})^2[1 + \rho e^{-b(y_i + y_{i+1})}].
\end{equation}
The single strand elasticity is represented by $k$, while the anharmonicity in the strand elasticity 
is represented by $\rho$. The parameter, $b$, describes the range of anharmonicity. For our studies, 
we choose model parameters $k$ = 0.015 eV\AA$^{-2}${} , $\rho = 50.0$, $b$ = 0.35 \AA$^{-1}${} for the current 
investigation \cite{Theo-PRE-2010}. In past it has been shown that value of $k$ and $\rho$ defines the 
sharpness in the transition from doble strand to single strand \cite{cocco,navin-epje05}. The hydrogen 
bonding between the two bases in the $i^{\rm th}$ pair is represented by the Morse potential as, 
\begin{equation}
\label{eqn2a}
V_M (y_i) = D_i(e^{-a_iy_i} - 1)^2. 
\end{equation}
where $D_i$ represents the potential depth, and $a_i$ represents the inverse of the width of the potential well. These two parameters have a crucial role in DNA denaturation. The dissociation energy, $D_i$, is a representation of the hydrogen bond energy that binds the $A−T$ and $G−C$ pairs while $a_i$ represents bond stiffness. From previous results, we know that the bond strengths of these two pairs are in an approximate ratio of $1.25$ - $1.5$ as the GC pairs have three while AT pairs have two hydrogen bonds\cite{ares,weber,zoli,zoli1,zoli2,ffalo,ffalo1,ffalo2,frank,macedo}. The potential parameters are taken as $a_{AT} = 4.2 \; {\rm \AA^{-1}},a_{GC} = 1.5\times a_{AT}$ and $D_{AT}$ = 0.075 eV while, $D_{GC} = 1.5\times D_{AT}$. 
\begin{table}[b]
\label{tab1}
\caption{Set of model parameters: The model parameters are tuned in such a way that the melting 
temperature of 120 base pairs chain in the free environment is $\sim$300 K. We find the $T_m^{free} = 315.38$ K with 
these parameters for the free chain.}
\begin{tabular}{|c|c|c|c|c|c|c|} \hline
$D_{\rm AT}$ & $D_{\rm GC}$ & $a_{\rm AT}$ & $a_{\rm GC}$ & $\rho$ & $\kappa$ & $b$ \\ 
(eV) & (eV) & (${\rm \AA^{-1}})$ & (${\rm \AA^{-1}}$) & (-)  & (${\rm eV/\AA^{2}}$) & (${\rm \AA^{-1}}$) \\ \hline
0.075 & $1.5D_{\rm AT}$ & 4.2 & $1.5a_{\rm AT}$ & 50.0 & 0.015 & 0.35 \\ \hline
\end{tabular}
\end{table}
We can study the thermodynamics of the transition by evaluating the partition function. For a sequence of $N$ base pairs, the canonical partition function can be written as \cite{navin-epje05}:
\begin{equation}
\label{eqn3}
Z = \int \prod_{i=1}^{N}\left\{dy_idp_i\exp(-\beta H)\right\} = Z_pZ_c,
\end{equation}
where $Z_p$ corresponds to the momentum part of the partition function while the $Z_c$ contributes as the configurational part of the partition function. The momentum part is nothing but $(2\pi mk_BT)^{N/2}$. The configurational part of the partition function, $Z_c$, is defined as \cite{navin-epje05}, 
\begin{equation}
\label{eqn4}
Z_c = \int \left[\prod_{i=1}^{N-1} dy_i  K(y_i,y_{i+1})\right]dy_NK(y_N)
\end{equation}
where 
$ K(y_i,y_{i+1}) = \exp\left[-\beta H_c(y_i,y_{i+1})\right]$. For the homogeneous chain, one can evaluate the partition function by transfer integral (TI) method by applying the periodic boundary condition \cite{zhang1997}. For the homogeneous chain, one can evaluate the partition function using the transfer integral (TI) method by applying the periodic boundary condition. For a chain with random sequence of $AT$ and $GC$ pairs and open boundaries, the calculation of partition function is little bit tricky. In past, various researchers have addressed the issue of solving the partition function of the chain with heterogeneous sequence and open boundaries \cite{cule1997,zhang1997,ns2011,campa1998,amar2013}. In addition to these issues, the partition function in the PBD model is divergent in nature. To overcome this problem, an upper cut-off for the integration needs to set up \cite{zhang1997,erp,pbd95,zhang1997,amar_pre15}. In our previous studies, we found that an integration limit of -5.0 ${\rm \AA}$ to 200 ${\rm \AA}$ is sufficient to overcome the divergence issue of the partition function. Once the proper cut-offs are introduced, the task is to discretise the integral in Eq.\ref{eqn4}. In order to get a precise value of melting temperature ($T_m$) we have observed that Gaussian quadrature is the most effective quadrature. We have found that discretization of the space with 900 points is sufficient to get an accurate value of $T_m$ \cite{amar_pre15,zhang1997}. Once we are able to evaluate the partition function, we can determine the thermodynamic quantities of interest by evaluating the Helmholtz free energy of the system. The Helmholtz free energy per base pair is defined as,
\begin{equation}
\label{eqn5}
f(T) = -\frac{1}{2}k_B T\ln\left(2\pi m k_B T\right) - \frac{k_B T}{N}\ln Z_c
\end{equation}
The specific heat, $C_v$, of the system, in the thermal ensemble, is evaluated by taking the second derivative of the free energy, as $C_v = -T(\partial^2 f/\partial T^2)$. The melting temperature ($T_m$) of the chain is evaluated from the peak in the specific heat. Another important quantity of interest is the average separation, $\langle y_j \rangle $, of the $j^{th}$ pair of the chain which is given by,
\begin{equation}
\langle y_j \rangle = \frac{1}{Z}\int \prod_{i=1}^N y_j \exp(-\beta H) dy_i
\end{equation}

\section{Introducing the Confinement in The Model}

\begin{figure}
\begin{center}
\includegraphics[height=1.3in,width=2.5in]{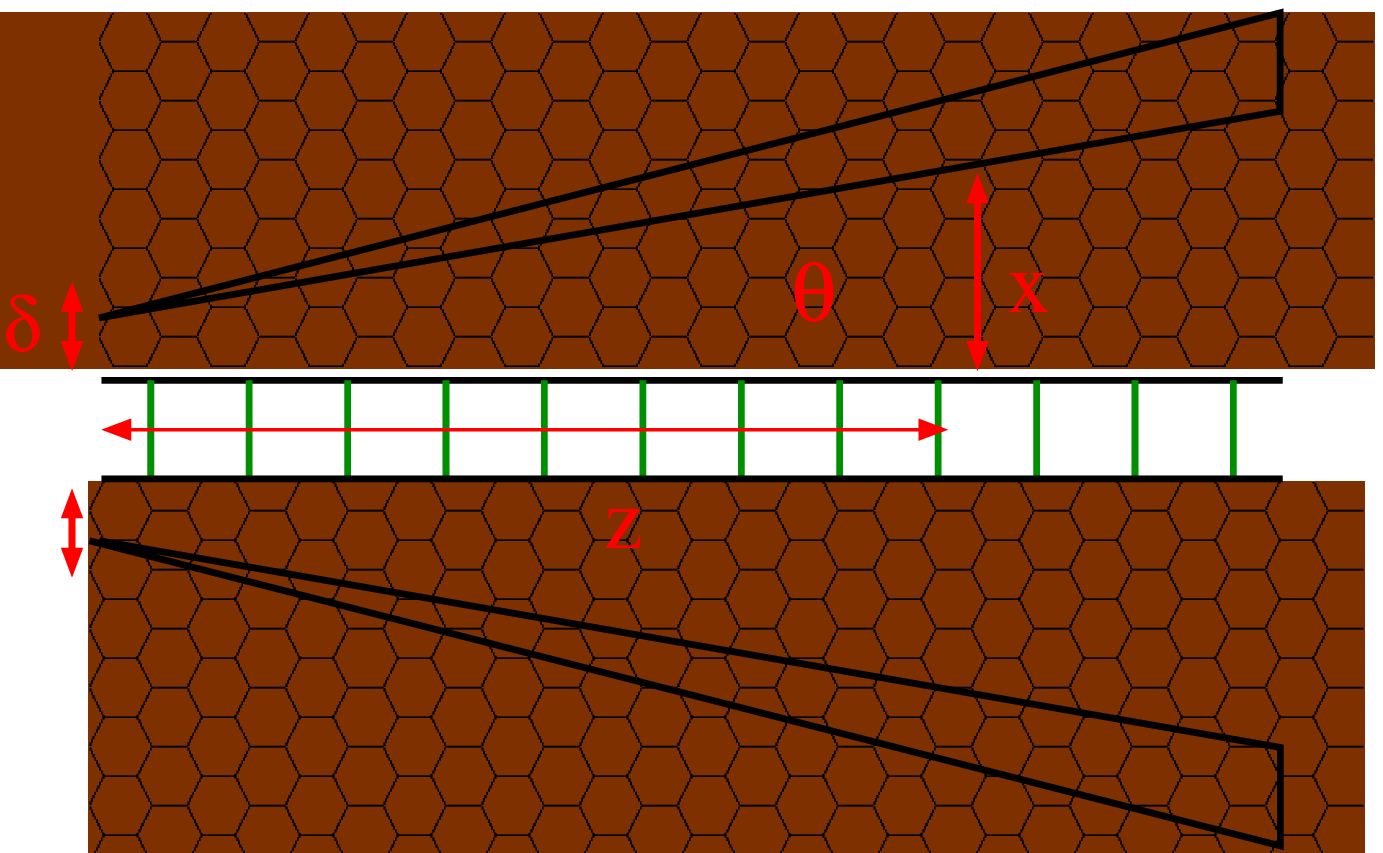}
\includegraphics[height=0.75in,width=2.5in]{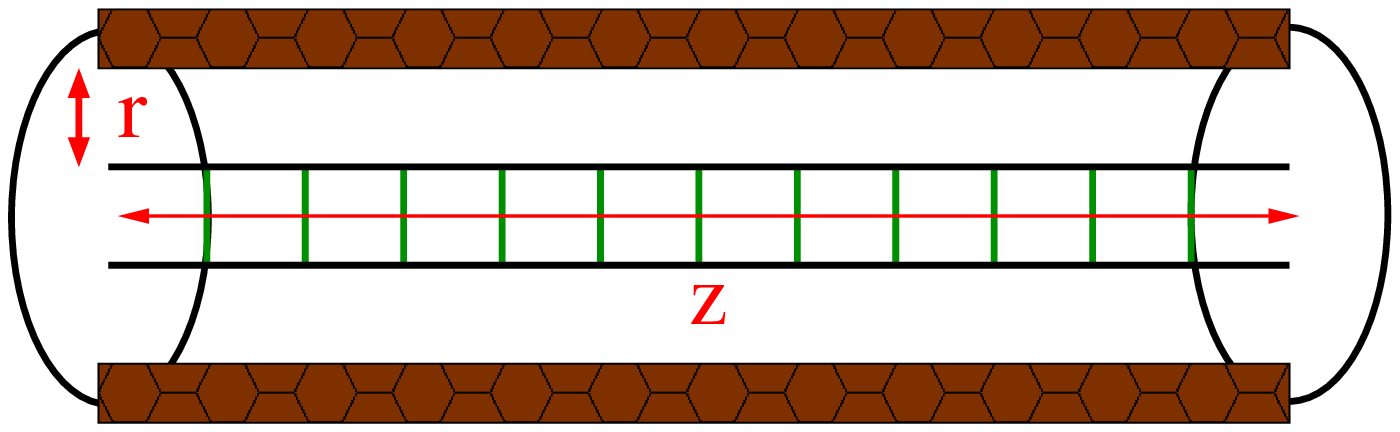}
\caption{\label{fig01} (a) The schematic representation of the DNA molecule in a confined cell of conical shape. The distance $x$ represents the distance between a pair and the confining wall while $\theta$ is the angle at which the cellular wall is suppose to confine the DNA. Here $z = 3.4\times i$, the distance along $x$-axis. (b) The schematic representation of the DNA molecule in a confined cell of cylindrical geometry. The $r$ is the distance of the confined wall from the DNA strand. The radius of the cylinder becomes $R_c$ = $r$ + DNA radius (10 ${\rm \AA}$).}
\end{center}
\end{figure}

In this work, we study the effect of confined geometries on the melting profile of $phage-\lambda$(GenBank: J02459.1) DNA chain of first 120 base pairs sequence. In past, researchers have studied the effect of rectangular confinement on the melting profile of the DNA molecules \cite{Mogurampelly,Michaela}. In this manuscript our focus is on effect of conical as well as cylindrical geometry (as shown in Fig.\ref{fig01}) on the thermal stability of the DNA molecule.
How the confinement can be realized in the PBD model? We adopt the following scheme. We restrict the configuration space of the system as shown in Fig. \ref{fig01}. There is no change in the lower limit of integral for both the geometry, however, the upper limit of integral 
for each base pair is modified according to the distinct geometry. For the conical confined geometry, the upper limit, $x$ is evalauted as, $x = 3.4\times i\times \tan(\theta)$ where $i$ is the site index ($i = 1,2,3,...,N$), and $N$ is the total number of base pairs in the chain. The term $\theta$ is the angular separation between the confining wall and the DNA strand. Since two base pairs are 
approximately 3.4 ${\rm \AA}$ distance apart we take this as the multiplication factor in calculation. Thus, the configurational space for our calculation extends from -5 ${\rm \AA}$ to $x \; {\rm \AA}$. There is another important factor that affects the melting transition is the cross section of the pore. The cross section of the confining wall (or pore width) is represented by the term, $\delta$. 
The value of $\delta = 0 \;{\rm \AA}$ means the distance between the vortex and the DNA strand is 0 ${\rm \AA}$ (as shown in Fig. \ref{fig01}a). For the cylindrical geometry the upper limit is defined as $r$. The quantity $r $ denotes the distance between the confined wall and the DNA strand. Using the modified scheme we calculate the partition function and hence evaluate all the thermodynamical properties of the system for both geometries.

\section{Results}
\label{results}
\begin{figure*}[hb]
\centering
\begin{center}
\includegraphics[height=2.5in,width=5.5in]{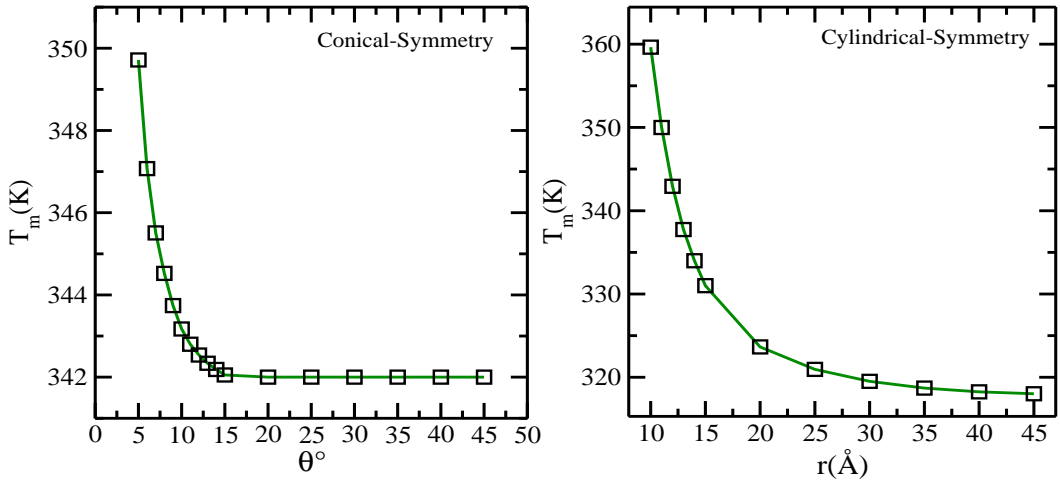}
\caption{\label{fig02}(on left)The change in the melting temperature with the change in the angular separation between the confining wall and DNA molecule. The pore width $\delta = 0 \; {\rm \AA}$. Figure shows that beyond a certain angle, $\theta_c$ there is no effect of confining wall on the melting temperature, $T_m$ of the molecule. (on right) The DNA is confined in a cylinder and the plot shows the change in melting temperature with radius of the cylinder.} 
\end{center}
\end{figure*}

We consider the first 120 bp of the phage-$\lambda$ DNA chain sequence and restrict it according to the confinement geometry. By evaluating the partition function, we calculate free energy per base pair hence specific heat as a function of temperature. The melting temperature $T_m$ of chain at various angular separations($\theta$) for conical confined geometry and in different diameter of the cylinder for cylindrical confined geometry are calculated through the peak in specific heat as a function of temperature \cite{zhang1997,Maity2017}. The results are plotted in Fig. \ref{fig02}. The obtained results clearly show that the confinement has a strong effect on the stability of the molecule. We find that the melting temperature ($T_m$) of dsDNA increases as the confinement becomes stronger. We vary the anglular separation and found that $T_m$ decreases as the angle changes from $5^0$ to $6^0$. The melting temperature for an angular separation of $5^0$ is found to be 349.7 K while it reduces to 347 K for an angle of $6^0$, {\it i.e.} a reduction of approximately $2.7$ K in the melting temperature. The melting temperature further reduces for the higher angular separations and get saturated at an angle of $\approx 15^0$ (342K). The stability of DNA confinement in a cylindrical symmetric wall has been investigated by various researchers using Poland Scheraga model \cite{Michaela,werner_pre_2015}. 

To investigate the effect of geometries on the stability of the DNA molecule, we also consider a DNA chain confined in a cylindrical confinement. We change the radius of the cylinder ($R_c$ = $r$ + DNA radius) and calculate the change in the melting temperature of the system. The results are plotted in Fig.\ref{fig02}. When the DNA is confined in a cylinder at $r = 10\; {\rm \AA}$, we find the melting temperature of the system as $\sim 359.6$ K. The melting temperature reduces drastically to a value of $\sim 350$ K for the radial separation of $11\; {\rm \AA}$(reduce to 9.6K). On further increasing the separation the $T_m$ reduces and almost get saturated at value of $\sim 318$ K. Since the two geometries supress the entropy of the system in a different way, one can believe that DNA which is confined in cylindrical geometry is more stable than the DNA that is confined in conical geometry upto a certain separation. The point to note that while for shorter separation stability is higher for cylindrical confinement but it is lower at higher separation ({\it please see Fig. \ref{fig02}} for a detail comparision).

\begin{figure}[H]
 \centering
\begin{center}
\includegraphics[height=2.75in,width=5.8in]{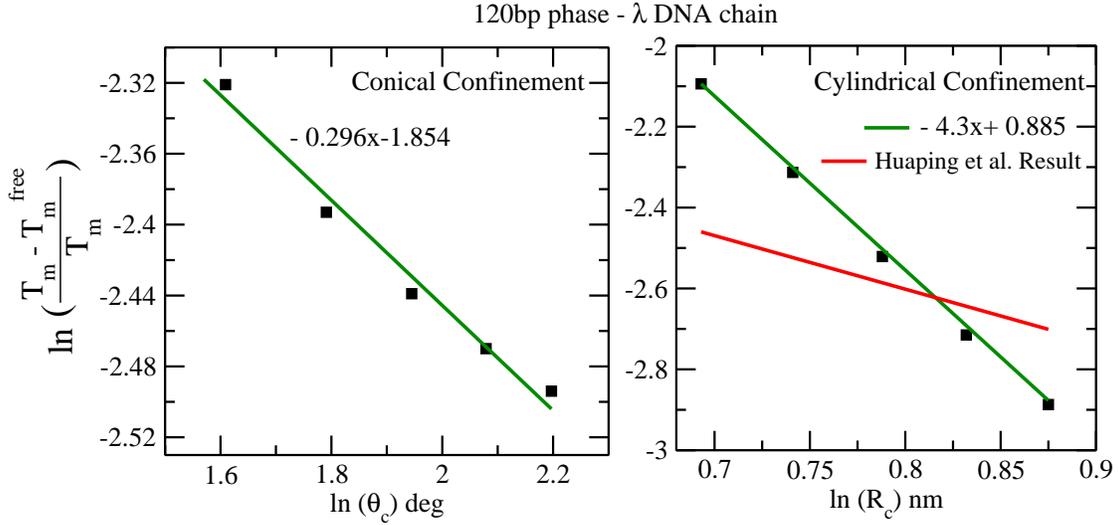}
\caption{\label{figboth}Relative $T_m$ shift follows as $\theta_c^{-v}$ and $R_c^{-v}$, where $v$ is 0.296 for the conical confinement and 4.3 for the cylindrical confinement. For the cylindrical confinement, the simulation result of reference \cite{Huaping_jcp14} is also plotted where $v$ is 1.323. The exponential index $v$ is achieved by fitting the straight lines.}
\end{center}
\end{figure}
We also compare our results with the results obtained by Huaping {\it et al} \cite{Huaping_jcp14}. They have calculated the relative change in melting temperature, $\frac{T_m - T_m^{free}}{T_m}$, with the radius of cyliner, $R_c$. With the help of numerical simulation, they found the  relative change in $T_m$ scales with the radius of cylinder as $R_c^{-v}$ with $v=1.323$. To compare our results, we also calculate the relative change in $T_m$ with the change in the radius of cylinder ($R_c$) and the angular separation of the conical confinement ($\theta_c$). The obtained results are plotted in the figure \ref{figboth}. Our results show that the relative change in $T_m$ scales with the radius of cylinder as $R_c^{-v}$ with $v=4.3$ for cylindrical confinement while $v = 0.296$ for conical confinement. Although the our system is little different in terms of sequence and chain length, our results are also showing similar power law behaviour with different exponent.  

To understand more about the microscopic details of opening of the DNA chain in a confined space, we calculate the average separation of the chain, $\langle y_i \rangle$, with the change in the angular separation ($\theta$). From the opening profile, the effect of confinement angle is more apparent. When the angular separation is $5^0$, the effect lasts for $\sim$25 base pairs while for the $10^0$ angular separation, the effect lasts for $\sim$ base pairs.
\begin{figure}
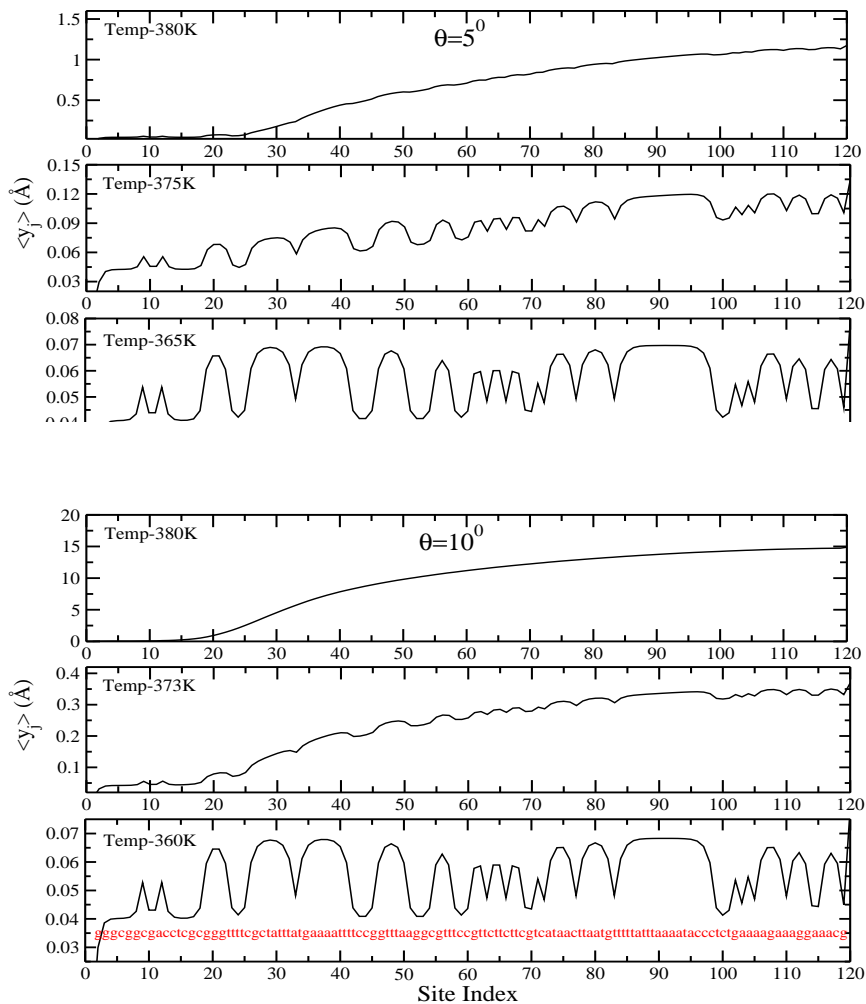

\begin{center}
\includegraphics[height=2.6in,width=4.5in]{diagrams/fig03.eps}
\includegraphics[height=2.6in,width=4.5in]{diagrams/fig04.eps}
\caption{\label{fig04} The opening of the chain in different angular separation($\theta$=$5^{0}$\& $10^{0}$). When the angular separation is $5^0$, the effect lasts for $\sim$25 base pairs while for the $10^0$ angular separation, the effect lasts for $\sim$20 base pairs at 380K.}
\end{center}
\end{figure}
How the different sections of the chain open below the melting temperature is also the matter of interest here. From the plot (fig. \ref{fig04}), the sequence and effect of conical confinement are visible. The weaker section of the chain consists of AT pairs is more entropic than the section containing GC pairs. While the opening of the homogeneous chain is very smooth, the opening of the heterogeneous chain has bubbles. One can see that how these bubbles disappear above the melting temperature of the chain when the DNA is in the denatured state.

\begin{figure}
\begin{center}
\includegraphics[height=2.3in,width=3.2in]{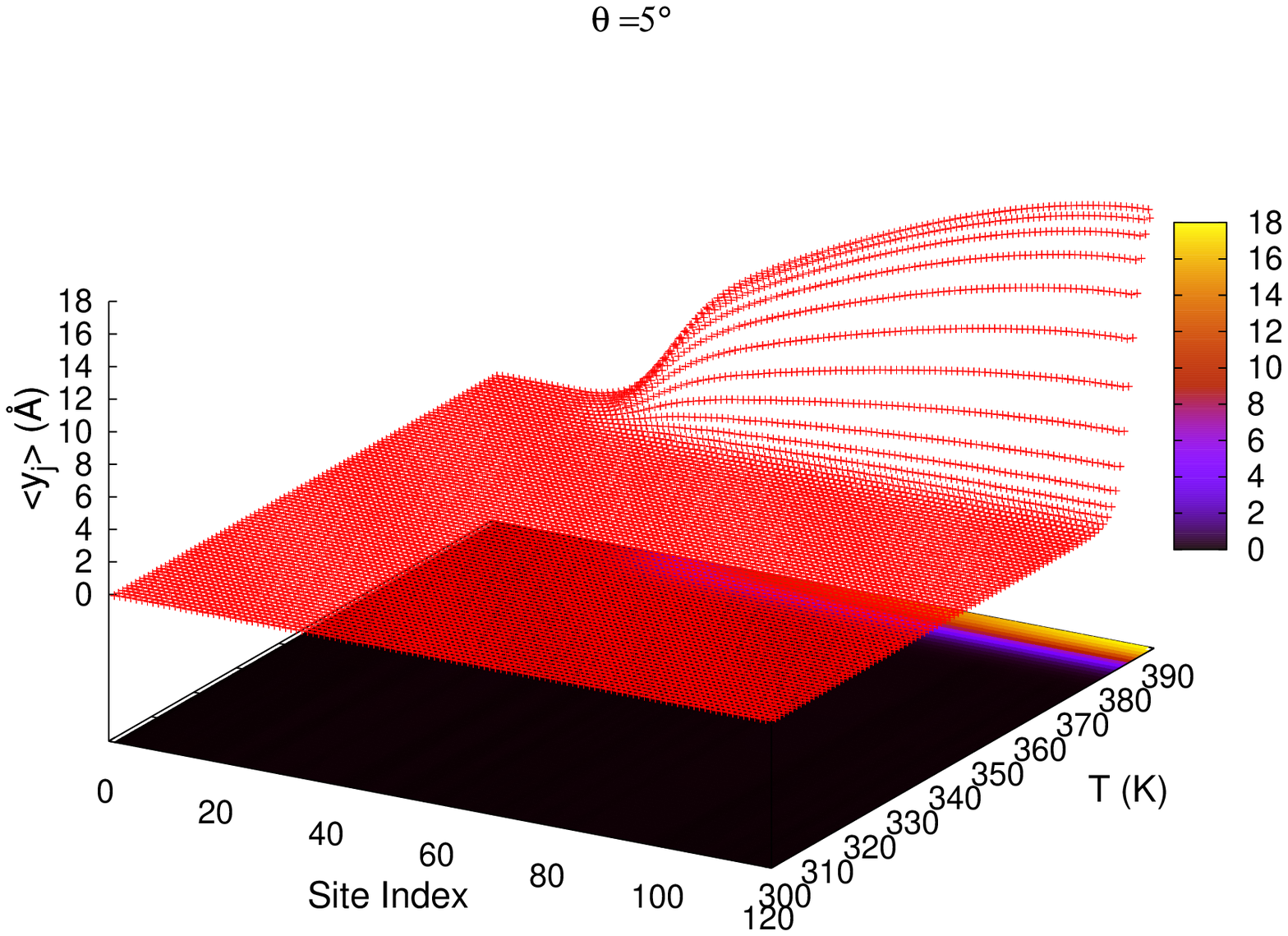}
\includegraphics[height=2.3in,width=3.2in]{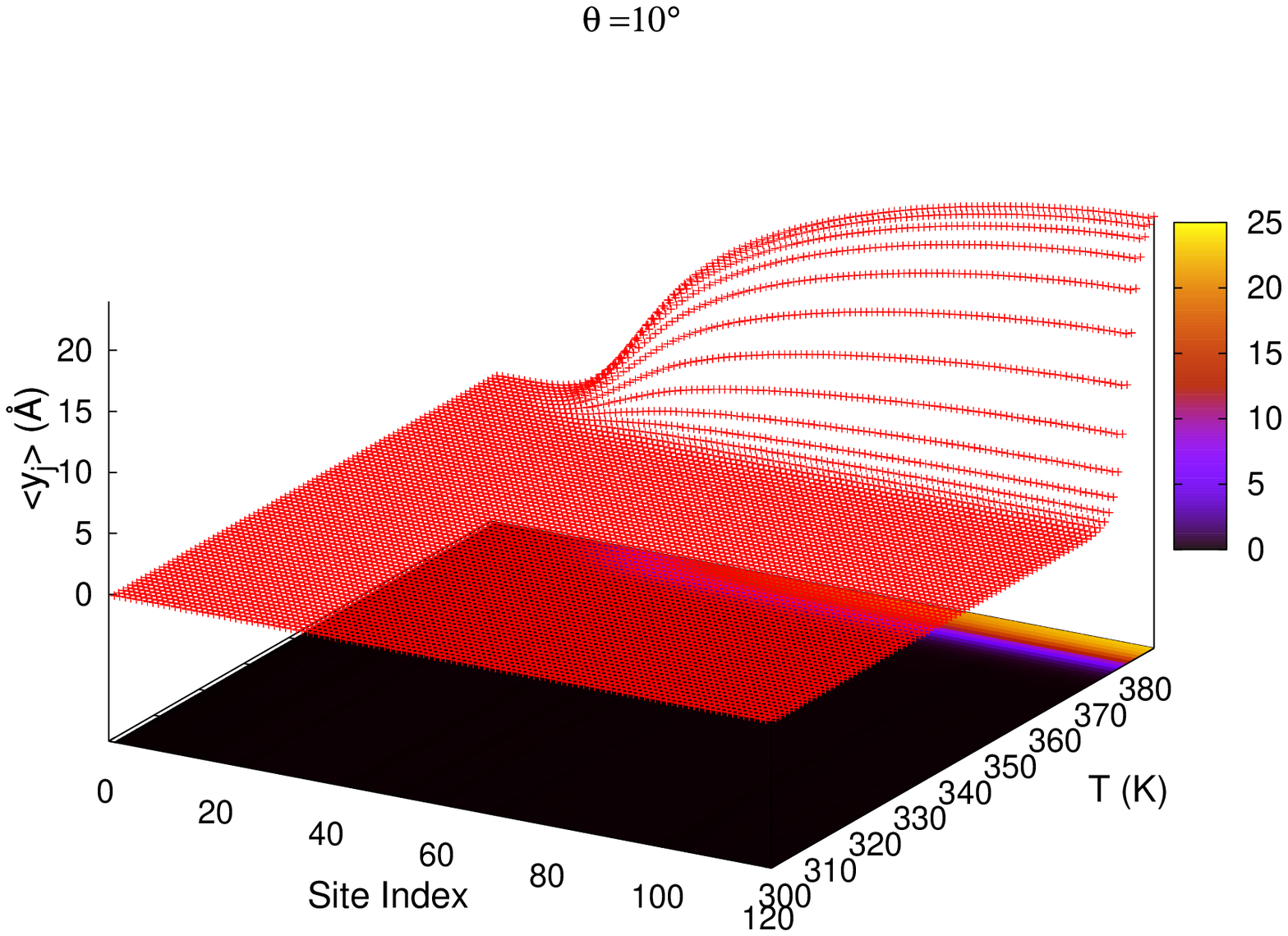} 
\includegraphics[height=2.3in,width=3.2in]{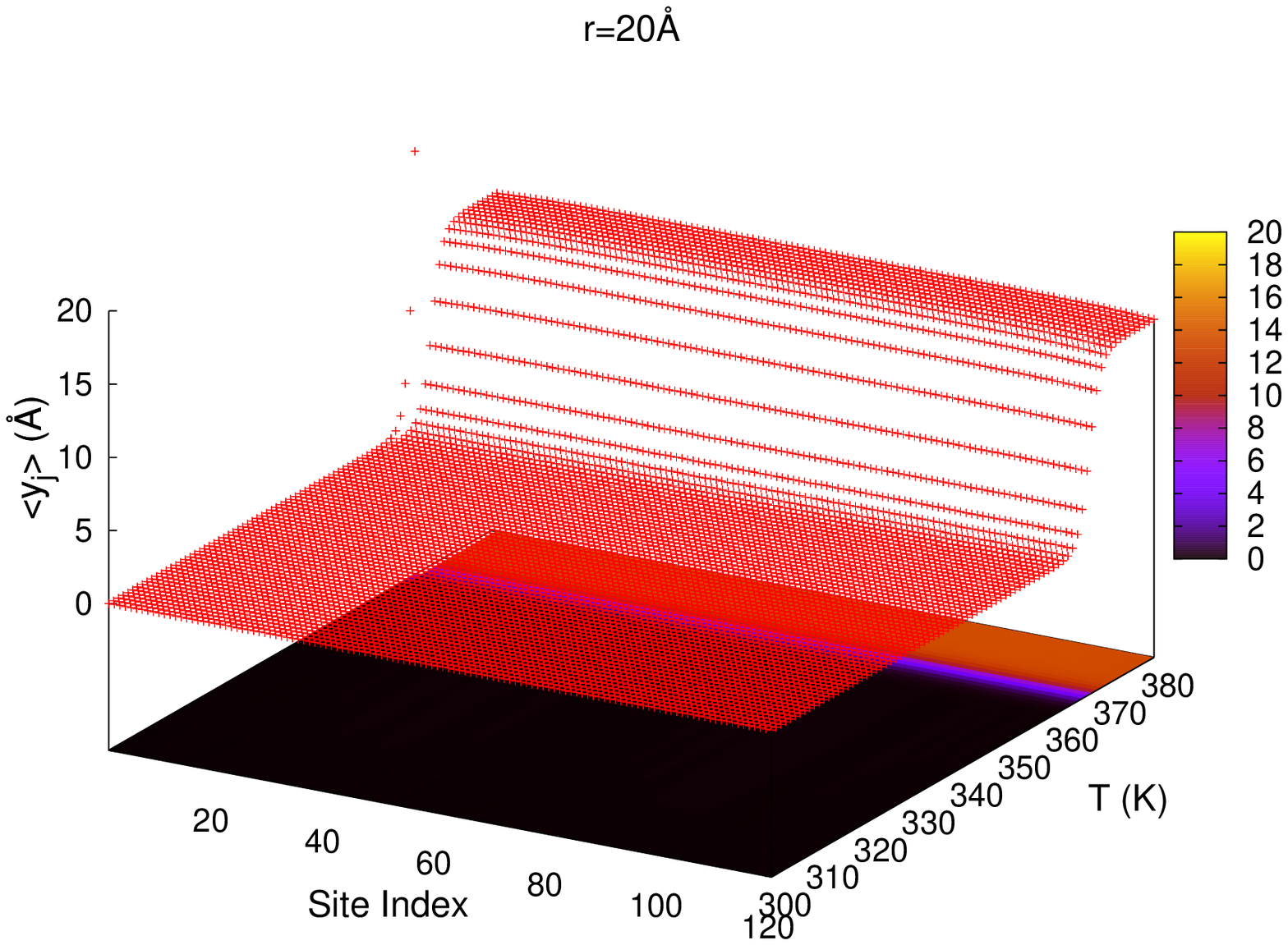}
\caption{\label{fig05}  The opening profile of the chain is shown through the 3D plots which shows a variation in $\langle y_j \rangle$ for the 120bp chain. Here we show the profile for two different angular separation, $\theta = 5^0$ and $\theta = 10^0$ (for pore width $\delta = 0 \; {\rm \AA}$) and for the cylindrical geometry $r = 20 \; {\rm \AA}$. The $x$-axis is site index while $y$-axis is temperature (in K) and $z$-axis denotes the average separation, $\langle y_j \rangle$ in ${\rm \AA}$.}
\end{center}
\end{figure}

Depending on the geometry the opening profile differs. The opening profile of the chain in these two different geometry are shown through 3D plots in Fig.\ref{fig05} which shows the variation of $\langle y_i \rangle$ with temperature and base pair.

\begin{figure}[hbt]
 \begin{center}
\includegraphics[height=3.5in,width=5.0in]{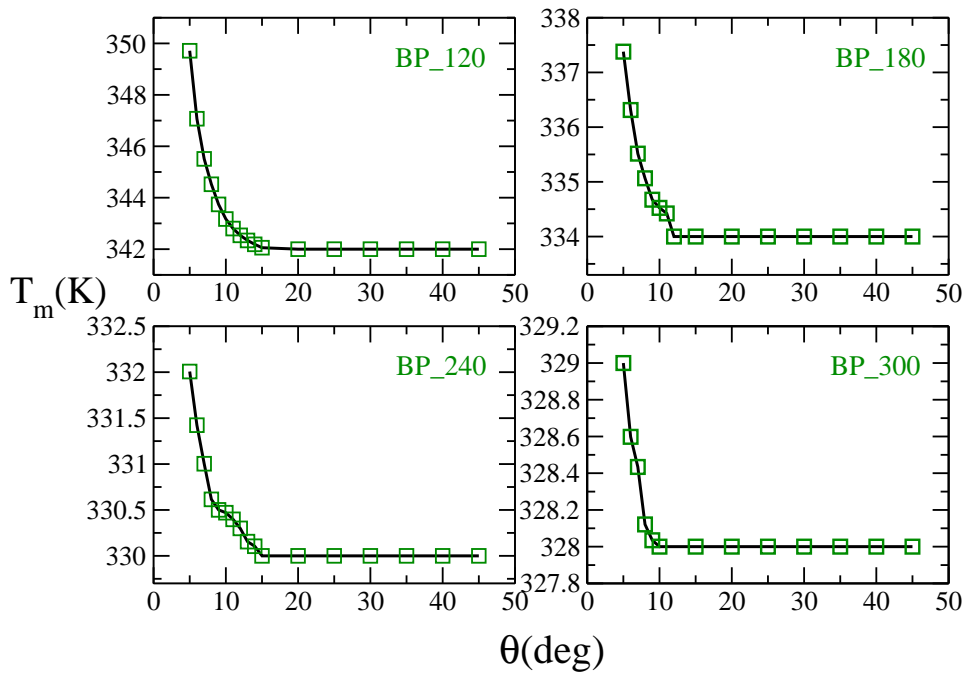}
\caption{\label{fig06} How the effect of confinement changes with the linear size of the molecule is shown here. We consider the phage-lambda DNA molecule sequence of different lengths as 120, 180, 240 and 300. The pore diameter $\delta = 0 \; {\rm \AA}$.}
\end{center}
\end{figure}
We also investigate the effect of confinement on different sized DNA molecule by varying the chain size. We consider the chain sequence of 120, 180, 240, and 300 base pairs of the phage-$\lambda$ DNA sequence and investigate the variation in melting temperature with varying angular separation. The resulting plots are shown in Fig. \ref{fig06}.The qualitative feature of all the plots are same, however, the decrease in melting temperature for all the chains is not same. For the chain of 120 bp, the $T_m$ is 349.7 K for an angular separation of $5^0$ which reduces to 347 K for and angle of $6^0$. The asympototic value of $T_m$ is 342 K which occur at $\approx 15^0$ of angular separation. This means for a span of $10^0$ ($15^0-5^0=10^0$) the variation in melting temperature vanishes and the decrease in $T_m$ is $\approx$5 K($349.7-342=7.7$ K). For the 180 bp chain,this shift is about 3.3 K ($337.3-334=3.3K$) for a span of $7^0$ ($12^0-5^0=7^0$) as it gets the asympototic value of $T_m$ at $12^0$, while, for 300 bp chain the melting temperature reduces from 329 K (at $5^0$) to 328 K (at $9^0$) so the shift of $T_m$ is $1$ K ($329-328 = 1$ K). One can clearly see that as the length of the chain increases the change in melting temperature with angle is reducing. For chain of length more than 350 base pairs, we observe that there is no change in the melting temperature with the change in the angular separation. The probable arugement for the small variation is the entropy of the system that varies with size. The smaller the chain, the stronger the suppression of entropy. For a bigger chain, a large section of the chain might not be seeing the confining wall because the distance between the wall and the DNA chain increases (please refer to Fig.01 a). 

\begin{figure}
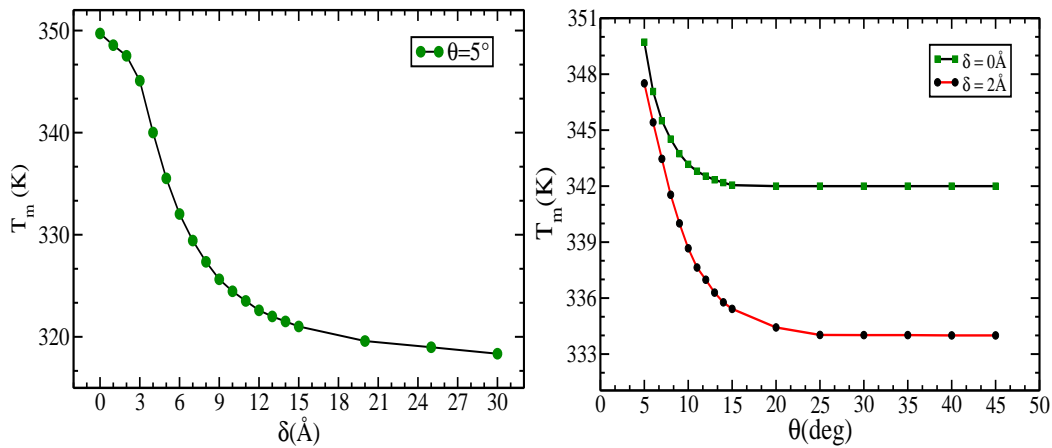

\includegraphics[height=2.3in,width=2.7in]{diagrams/fig09.eps}
\includegraphics[height=2.3in,width=2.7in]{diagrams/fig10.eps}
\caption{\label{fig07} (On left) The change in the melting temperature ($T_m$) with the pore width ($\delta$) for a fix value of angular separation ($\theta = 5^0$). (On right) The change in melting temperature ($T_m$) with the confinement angle ($\theta$) for two values of the pore width ($\delta$).}
\end{figure}

Next, we study the thermal stability of the system with the change in the pore width,$\delta$ for conical geometry. Here we calculate the change in the melting temperature with the change in the pore width, $\delta$ for a fix value of $\theta$ (the angular separation). The results are shown in Fig. \ref{fig07}.
From the Fig. \ref{fig07} it is clear that with the increase in pore width ($\delta$) for a same value of angular separation ($\theta$), the melting temperature decreases. This is due to fact that when the pore width increases the end at the vortex of the cone gets more space which increases the end entropies. While for the lower pore diameter, only one of the ends of the DNA has sufficient entropy to contribute to the opening of DNA molecule, for higher values of pore diameter, the other end also contribute to the opening of the molecule. Thus, the DNA molecule now melts at lower temperature. 

\section{Conclusions}
\label{conc}

In this manuscript, we have studied, theoretically, the stability of phage-$\lambda$ DNA molecule of different lengths that are confined in conical as well as cylindrical geometry. Probably this is the first model-based study that studies the impact of angular confinement on the stability of the DNA molecule in a thermal ensemble. Our studies show that not only the confinement but also the geometry of the confinement plays a crucial role in the stability as well as the overall activity of the DNA molecule. 

When the confining wall becomes closer, the configuration space for the DNA molecule, in cylindrical geometry, is restricted, which suppresses the entropy of the chain. Hence the melting temperature of the system in cylindrical geometry is more than the melting temperature of the DNA that is confined in conical geometry up to a certain extends. If we compare two cases of the confinement (conical and cylindrical) and find the rate of decrease in the $T_m$, then we see that the DNA confined in cylindrical geometry has a higher decrease rate. Through simulation of the short DNA molecule that is confined in a cylindrical as well as in spherical geometry, Huaping {\it et al.} \cite{Huaping_jcp14} has found that the melting temperature of the system decreases with the increasing diameter of the cylinder/sphere. We have found through the PBD model that the relative change in $T_m$ scales with the radius of cylinder as $R_c^{-v}$ with $v=4.3$ for cylindrical confinement while $v = 0.296$ for conical confinement. Since our system is little different in terms of sequence and chain length than the Huaping {\it et al}, our results are also showing similar power law behaviour with different exponent. 

Our findings may attract experimentalists to design the shells of different geometries and investigate the effect of the geometry of the shell on the stability of encapsulated DNA. In DNA encapsulation, the ability to preserve the DNA and efficiently release it, are the processes which are inversely related. The objective of good DNA encapsulation is to find an optimal balance between these two issues. Researchers are trying to balance these two issues to find a potential material for DNA encapsulation.

\section*{Acknowledgement}
We would like to acknowledge M. Peyrard (Lyon), Amar Singh (CCB, University of Kansas, USA), and Y. Singh (BHU) for useful discussions and drawing our attention to some useful papers. We appreciate the financial support from Department of Science and Technology, New Delhi.

\bibliography{ms_confined}

\end{document}